# RECIPROCAL SYMMETRY AND EQUIVALENCE BETWEEN RELATIVISTIC AND QUANTUM MECHANICAL CONCEPTS

Mushfiq Ahmad

Department of Physics, Rajshahi University, Rajshahi, Bangladesh

E-mail: mushfiqahmad@ru.ac.bd

## ABSTRACT

We have defined slowness (or reciprocal velocity) corresponding to *v* as $cc/v$, where *c* is the speed of light and *v* is the corresponding velocity. Velocity and slowness are images of each other. Reciprocal symmetric law of addition of velocities fulfils the requirement that the sum (or the difference) of velocities remains unchanged if velocities are replaced by the corresponding slownesses. We have postulated reciprocal symmetry, which states that *every valid statement has an equally valid (reciprocal) image statement.* The postulate has allowed us to derive:

1. Einstein's law of addition of velocities from Einstein's image postulate – *No physical body can travel with a slowness less than the slowness of light.*

2. Planck's Radiation Formula from Planck's image postulate – *No body can attain a level of energy higher than an upper limit value 2W.*

3. Reciprocal image of Boltzmann's probability function gives the Fermi-Dirac analogue of Planck's radiation formula

4. Heisenberg-like relation $E.t \geq cT^2$ from the postulate – *The highest attainable rate of energy transfer is c.*

5. Image function of classical harmonic oscillator exponential function gives quantum-like half-integral-multiple values of oscillator energy.

6. Image function of classical harmonic oscillator exponential function also gives energy levels of a free particle oscillating between rigid walls.

7. Reciprocal symmetry ensures that the speed of light is invariant under addition.

8. The general law of addition is given so that the same law of addition gives (depending on a reciprocity parameter) the relative velocity as well as the relative slowness.

**INTRODUCTION**

In 1900 Max Planck presented his quantum hypothesis[1], which sets a lower limit to energy transfer. In 1905 Albert Einstein presented his relativistic postulate[2], which sets an upper limit to velocity. Apart from the fact that, they talk of different quantities (energy and velocity), there is a reciprocal relation between the two.

Every number has a unique reciprocal (A discussion on the reciprocal of 0 is given is deferred to a later paper). There will be no ambiguity if, instead of representing distance, time etc. by $x, t$ etc., we represent them by their reciprocals ($x \rightarrow 1/x$, $t \rightarrow 1/t$ etc.). Numerical value of the speed, with this choice, will be the reciprocal of the usual value[3]. Therefore, the kinematical description of an event in terms of reciprocal values should be as valid as a description in terms of usual values. We shall adopt this viewpoint. A physical theory should be objective and should not depend upon the attitude of the observer.

Space and time are measured quantities. Speed and slowness (Speed = space/time and slowness = time/space) are quantities an observer defines. A physical theory should be objective and physics is independent of definitions. Therefore, the kinematical description of an event in terms of slowness should be as valid as a description in terms of speed.

**SLOWNESS AND RECIPROCAL SYMMETRY**

*A physical theory should be such that it permits both direct and reciprocal representations.*

Example: Relative velocity should remain invariant if velocities are replaced by corresponding slownesses. e.g.

$$u \oplus (\pm v) = \frac{u \pm v}{1 \pm u.v} = u^* \oplus (\pm v^*) \tag{1}$$

Where

$$v^* = 1/v \tag{2a}$$

**Image of Einstein's Postulate**
(Einstein's postulate in Planck's Form)

We translate Einstein's postulate as below:

*No body can move with a slowness, $v^*$, less than 1.*
Substituting $v \to v/c$ and $v^* \to v^*/c$ we get

$$v^* = c^2/v \tag{2b}$$

Einstein's postulate now reads:

*No body can move with a slowness, $v^*$, less than $c^*=c$, the slowness of light.*

$$v^* \geq c \tag{3}$$

**Slownesses are Discrete**

The difference between two slownesses is also a slowness. It is, therefore, greater than (or equal to) *c*. Therefore, slownesses are discrete.

**Slownesses of the Body at Rest**

By (2), the slowness corresponding to velocity 0 is $0* = \dfrac{c^2}{0}$

We shall follow the prescription

$$0* = Lt_{\varepsilon \to 0} \dfrac{c^2}{\varepsilon} \tag{4}$$

**Law of Addition of Slownesses**

The sum or the difference of two slownesses should be a slowness and, therefore, must fulfill Einstein's postulate (in Planck's form)

If $w*$ is the sum of slownesses $u*$ and $v*$, $w*$ is given by

The sum

$$w* = (u*) \oplus *(v*) = \dfrac{(u*+1)(v*+1)+(u*-1)(v*-1)}{(u*+1)(v*+1)-(u*-1)(v*-1)} \tag{5}$$

Or, introducing $c$ ($w* \to w*/c$) etc.

$$w* = \dfrac{c^2 + u*.v*}{u*+v*} \tag{6}$$

The postulate is satisfied

$$c\{u* \oplus *(\pm v*)\} \geq c \ for \ |u*|,|v*| \geq c \tag{7}$$

**Einstein's Law of Addition of Velocities**

Einstein's image postulate gives Einstein's law of addition of velocities.

Using relation (2b) we write relation (6) in terms of velocities $u$, $v$ and $w$. This gives

$$w = c\,(u \oplus v) = \frac{u+v}{1+u.v/c^2} \tag{8}$$

We have introduced the notation

$$u \oplus v = \frac{c^2}{u \oplus * v} \tag{9}$$

(8) is Einstein's law of addition of velocities.

In deriving (8) we have used the symmetry relation

$$(u^*) \oplus * (v^*) = u \oplus * v \tag{10}$$

## RECIPROCAL SYMMETRY AND INVARIANCE OF *c* UNDER ADDITION

A consequence of reciprocal symmetry is that replacing one of the terms (of a sum) by its reciprocal replaces the whole sum by its reciprocal.

$$u \oplus (c^2/v) = \frac{c^2}{u \oplus v} \tag{11}$$

$v=c$ in (11) gives

$$u \oplus c = \frac{c^2}{u \oplus c} \tag{12}$$

Therefore

$$u \oplus c = \pm c \tag{13}$$

(13) should remain valid for $u=0$. Therefore

$$u \oplus c = c \tag{14}$$

## PLANCK'S RADIATION FORMULA

The formula for the distribution of light in a black body is given by[4]

$$I(w)dw = <E> \frac{w^2 dw}{\pi^2 c^2} \tag{15}$$

with

$$<E> = \frac{\hbar w}{\exp(\hbar w/kT) - 1} \tag{16}$$

Planck's hypothesis has been invoked to find $<E>$.

The number $N_n$ of particles in nth energy level is given by Boltzmann's probability distribution relation[5]

$$N_n = N_0 \exp(-E_n . s) \tag{17}$$

In (17) will correspond to (16) $E_n = n\hbar w$ and $s = 1/kT$.

**Symmetric Boltzmann Equation**

$N_n$ is the solution of Boltzmann's equation[6]

$$\frac{dN_n}{ds} = -E_n N_n \tag{18}$$

To invoke symmetry principle we replace the above differential equation by the corresponding reflection symmetric finite difference equation below. Omitting the subscripts, corresponding to (18) we write

$$\frac{DN}{D(s,\delta)} = -\overline{E}N \tag{19}$$

where

$$\frac{DN}{D(s,\delta)} = \frac{N(E, s+\delta) - N(E, s-\delta)}{2\delta} \tag{20}$$

and

$$\delta = 1/W \tag{21}$$

We have called the (20) reflection symmetric because it remains invariant under the change $\delta \to -\delta$. In the limit $\delta \to 0$ we should get

$$\frac{DN}{D(s,\delta)} = -\overline{E}N \xrightarrow{\delta \to 0} \frac{dN}{ds} = -EN \tag{22}$$

**Image of Planck's Hypothesis**
(Planck's Hypothesis in Einstein Form)

*Postulate: No energy level can be higher than 2W or lower than -2W*

$$-2W \leq E \leq 2W \tag{23}$$

The factor 2 in $2W$ has been included for future algebraic convenience.
We have introduced the factor 2 for future algebraic convenience.

2 functions $N = f_1$ and $N = f_2$ satisfy the equation if

$$f_1 = A \left( \frac{1 - \frac{E}{2W}}{1 + \frac{E}{2W}} \right)^{W.s} \tag{24}$$

and

$$f_2 = A \left( \frac{1 + \frac{2}{E.\delta}}{1 - \frac{2}{E.\delta}} \right)^{s/\delta} = A \left( -\frac{1 + \frac{E.\delta}{2}}{1 - \frac{E.\delta}{2}} \right)^{s/\delta} \tag{25}$$

with

$$\overline{E} = \frac{E}{1 - (E/2W)^2} \tag{26}$$

To go from $f_1$ to $f_2$, we have taken the arguments to their reciprocals. $\frac{E}{2W} \rightarrow -\frac{2W}{E}$. In the Galilean limit $W \rightarrow \infty$, we get the classical function

$$f_1 = A \left( \frac{1 - \frac{E}{2W}}{1 + \frac{E}{2W}} \right)^{W.s} \xrightarrow[W \to \infty]{} A\exp(-E.s) \tag{27}$$

**Planck's Formula**

If we use $f_1$ instead of $\exp(-\hbar w / kT)$ in calculating $<E>$ and in the end we replace $f_1$ by its classical limit, we get Planck's result.

$$I(w)dw = <E> \frac{w^2 dw}{\pi^2 c^2} \tag{28}$$

with

$$<E> = \frac{E_1}{\exp(E_1.s) - 1} = \frac{\hbar w}{\exp(\hbar w/kT) - 1} \tag{29}$$

If we try to use $f_2$ instead of $f_1$ in calculating $<E>$, the function explodes as $n \to \infty$ for $E_n > 0$. Therefore, we have to restrict ourselves to the case of negative energies $E_n < 0$. When $n$ is odd, we can take the negative sign out. Therefore, for $n = 1$ the average energy becomes

$$<E> = \frac{\hbar w}{-\exp(-\hbar w/kT) - 1} = \frac{\hbar w'}{\exp(\hbar w'/kT) + 1} \tag{30}$$

where $\hbar w' = -\hbar w$

In this case Planck's formula becomes

$$I(w)dw = <E> \frac{w^2 dw}{\pi^2 c^2} \quad \text{with} \quad <E> = \frac{\hbar w'}{\exp(\hbar w'/kT) + 1} \tag{31}$$

For $n = even$, the above relation becomes

$$<E> = \frac{\hbar w}{\exp(-\hbar w/kT) - 1} = \frac{-\hbar w'}{\exp(\hbar w'/kT) - 1} \tag{32}$$

In this case $<E>$ is negative and Planck's formula becomes

$$I(w)dw = <E> \frac{w^2 dw}{\pi^2 c^2} \quad \text{with} \quad <E> = \frac{-\hbar w'}{\exp(\hbar w'/kT) - 1} \tag{33}$$

## HEISENBERG'S RELATION FROM PLANCK'S IMAGE POSTULATE

*Postulate: c is the upper limit of the rate of energy transfer*

If $E$ is energy transferred in time $t$

$$-c \leq E/t \leq c \tag{34}$$

**Reciprocal Symmetric Law of Multiplication**

The law of multiplication consistent with -- distributive with respect to – (8) is

$$z = n \otimes u = \frac{\left(\dfrac{1+u}{1-u}\right)^n - 1}{\left(\dfrac{1+u}{1-u}\right)^n + 1} \tag{35}$$

Substituting $z \to z/c, u \to u/c$

$$z = c(n \otimes u) = c\frac{\left(\dfrac{c+u}{c-u}\right)^n - 1}{\left(\dfrac{c+u}{c-u}\right)^n + 1} \tag{35}$$

so that

$$\begin{aligned}(n \otimes u) \oplus (m \otimes u) &= (n+m) \oplus u \\ (n \otimes u) \oplus (n \otimes v) &= n \otimes (u \oplus v)\end{aligned} \tag{36}$$

Similarly, the law of multiplication consistent with -- distributive with respect to -- (1) is, (including $c$).

$$z^* = c(n \otimes {}^*u) = c\frac{\left(\dfrac{u+c}{u-c}\right)^n + 1}{\left(\dfrac{u+c}{u-c}\right)^n - 1} = \frac{c^2}{n \otimes u^*} \tag{37}$$

so that

$$\begin{aligned}(n \otimes {}^*u^*) \oplus {}^*(m \otimes {}^*u^*) &= (n+m) \oplus {}^*u^* \\ (n \otimes {}^*u^*) \oplus {}^*(n \otimes {}^*v^*) &= n \otimes {}^*(u^* \oplus {}^*v^*)\end{aligned} \tag{38}$$

Values in the limit are

$$n \otimes u \xrightarrow[u/c \to 0]{} n.u \tag{39}$$

$$n \otimes *u* \xrightarrow[c/u^* \to 0]{} u*/n \qquad (40)$$

**Definition of Rate of Transfer**

Let $E$ be the energy transferred in time $t$. The rate of transfer is by (6)

$$y = (T/t) \otimes (E/T) = c \frac{\left(\frac{c+E/T}{c-E/T}\right)^{T/t} - 1}{\left(\frac{c+E/T}{c-E/T}\right)^{T/t} + 1} \leq c \qquad (41)$$

Where $T$ is a quantity having the dimension of time.

In the limit

$$y = (T/t) \otimes (E/T) \xrightarrow[E/(cT) \to 0]{} E/t \leq c \qquad (42)$$

The condition

$$E/t \leq c \qquad (43)$$

(43) is reciprocal symmetric form (Einstein's form) of Planck's hypothesis.

The rate of energy transfer, in reciprocal space, is by (8)

$$y^* = (T/t) \otimes *(E/T) = c \frac{\left(\frac{E/T+c}{E/T-c}\right)^{T/t} + 1}{\left(\frac{E/T+c}{E/T-c}\right)^{T/t} - 1} \geq c \qquad (44)$$

In the limit

$$(T/t) \otimes *(E/T) \xrightarrow[(cT)/E \to 0]{} Et/T^2 \geq c \qquad (45)$$

Therefore,

$$Et \geq cT^2 \qquad (46)$$

(46) is Heisenberg's image of (43).

## HARMONIC OSCILLATOR AND POTENTIAL WELL

Simple harmonic oscillator function $f$ satisfies the differential equation

$$\frac{df}{dt} = \pm iwf \tag{47}$$

The corresponding finite difference symmetric equation is

$$\frac{Dg}{D(t,\delta)} = \pm iwg \tag{48}$$

It has two solutions $g = g_1$ and $g = g_2$

$$g_1 = A \left( \frac{1 - i\frac{1-\sqrt{1+(iw\delta)^2}}{w\delta}}{1 + i\frac{1-\sqrt{1+(iw\delta)^2}}{w\delta}} \right)^{t/\delta} \tag{49}$$

$$g_2 = A \left( -\frac{1 + i\frac{1-\sqrt{1+(iw\delta)^2}}{w\delta}}{1 - i\frac{1-\sqrt{1+(iw\delta)^2}}{w\delta}} \right)^{t/\delta} \tag{50}$$

In the limit as $\delta \to 0$, $g_1$ gives the classical oscillator function $f$.

$$g_1 = A\, e^{2n\pi i} \left( \frac{1 \pm i\frac{1-\sqrt{1+(iw\delta)^2}}{w\delta}}{1 \mp i\frac{1-\sqrt{1+(iw\delta)^2}}{w\delta}} \right)^{t/\delta} \xrightarrow[\delta \to 0]{} A\exp(2n\pi/\delta \pm w)it \tag{51}$$

**Half Integral Energy Levels**

$g_2$ gives

$$g_2 = A \left( -\frac{1 + i\frac{1-\sqrt{1+(iw\delta)^2}}{w\delta}}{1 - i\frac{1-\sqrt{1+(iw\delta)^2}}{w\delta}} \right)^{t/\delta} \xrightarrow[\delta \to 0]{} A\exp iw't \tag{52}$$

where
$$w' = (2n+1)\pi/\delta + w = y + w \quad (53)$$

The energy of the oscillator is proportional to
$$(w')^2 = y^2 + 2yw + w^2 = \{(2n+1)\pi/\delta\}^2 + 2\{2(n+1/2)\pi/\delta\}w + w^2 \quad (54)$$

The middle terms contain half-integral multiples. To this extent it corresponds to quantum mechanical value.

**Energy Levels of a Free Particle Oscillating Between Reflecting Walls**

The first term of (54), $y^2 = \{(2n+1)\pi/\delta\}^2$, corresponds to the energy levels of a free particle oscillating between reflecting walls A similar term with $(2n+1)$ replaced by $2n$ comes from $g_1$ solution. Adding terms, we get the total energy, $E_y$, of a free particle oscillating between reflecting walls as

$$E_y = \frac{1}{2}m\{An\pi/\delta\}^2 \quad (55)$$

where $m$ is the mass, $2A$ is the width of the well and $n$ is an integer.

**Correspondence between Harmonic Oscillator and Potential Well**

(54) above contains term corresponding to both square well and oscillator energy levels. In the limit $w \to 0$, we are left with the square well part corresponding to (55) only.

**RECIPROCAL SYMMETRIC LAW OF ADDITION OF VECTORS**

Let **U** be the velocity of a body and **V** be the velocity of the observer. The reciprocal symmetric relative velocity (difference of velocities) **W** is defined as[7]

$$\mathbf{W} = \mathbf{U} \mathbin{\hat{+}} (-\mathbf{V}) = \frac{\mathbf{U} - \mathbf{V} - i\dfrac{\mathbf{U} \times \mathbf{V}}{c^2}}{1 - \dfrac{\mathbf{U} \cdot \mathbf{V}}{c^2}} \quad (56)$$

**Rotation in Reciprocity Space**

$\widetilde{\mathbf{W}}$, which is $\mathbf{W}$ rotated in rotation space through angle $\phi$, in an arbitrary direction defined by the unit vector $\mathbf{r}$, is defined by the following rotation operation

$$\widetilde{\mathbf{W}} = \frac{\mathbf{W} + \left(i\mathbf{r} - \dfrac{\mathbf{W} \times \mathbf{r}}{c^2}\right)\tan(\phi/2)}{1 + i\dfrac{\mathbf{W} \cdot \mathbf{r}}{c^2}\tan(\phi/2)} \quad (57)$$

When $\phi = 0$ there is no rotation. When $\phi = \pi$, we get

$$\widetilde{\mathbf{W}} \xrightarrow[\phi \to \pi]{} \frac{\left(i\mathbf{r} - \dfrac{\mathbf{W} \times \mathbf{r}}{c^2}\right)}{i\dfrac{\mathbf{W} \cdot \mathbf{r}}{c^2}} \quad (58)$$

In this case $\widetilde{\mathbf{W}}$ is the reciprocal of $\mathbf{W}$ in the sense that

$$\mathbf{W} \cdot \widetilde{\mathbf{W}} \xrightarrow[\phi \to \pi]{} c^2 \quad (59)$$

**Reciprocal Symmetry**

We now have reciprocal symmetry

$$\widetilde{\mathbf{U}} \mathbin{\hat{+}} (-\widetilde{\mathbf{V}}) = \mathbf{U} \mathbin{\hat{+}} (-\mathbf{V}) \quad (60)$$

**RECIPROCITY-GENERAL ADDITION OF VELOCITIES**

The general (covering both velocity and reciprocal velocity) relative velocity/slowness is

$$\tilde{W} = \frac{\tilde{U} - \tilde{V} - i\frac{\tilde{U} \times \tilde{V}}{c^2}}{1 - \frac{\tilde{U}.\tilde{V}}{c^2}} \qquad (61)$$

gives the general law of relative velocity/slowness.

$\phi = 0$ gives the relative velocity and $\phi = \pi$ gives the relative slowness.

When $\phi = \pi$ and $\mathbf{r} = \frac{\mathbf{W}}{|\mathbf{W}|}$

$$\tilde{W} = \frac{W}{W.W} \qquad (62)$$

**ASSOCIATIVITY**

We can verify that[8]

$$(U \hat{+} V) \hat{+} Y = U \hat{+} (V \hat{+} Y) \qquad (63)$$

for any **U**, **V** or **Y**.

**CONCLUSION**

We have been able to establish equivalence between quantum mechanical and relativistic concepts by deriving relativistic results from quantum mechanical postulates (like Einstein's law of addition of velocities from the postulate that slowness has a lower limit), and vise-versa (like Planck's law from the postulate that energy has an upper limit).

Using reciprocal symmetric symmetry and classical consideration only, we have been able to derive half-integral values of energy of a simple harmonic oscillator and square integral values of energy of a particle oscillating between reflecting walls. No quantum mechanical postulates or Planck's hypothesis are involved.

We have also derived the general law of addition of velocities so that the same law gives (for the different values of rotation parameter), the relative velocities as well as the relative slowness. We have also shown that reciprocal symmetric law of addition is associative.

---

[1] Max Planck. Annalen der Physik 4 (1901): 553.
[2] On the Electrodynamics of Moving Bodies. *Annalen Der Physik*, 17, 1905
[3] "The question on the reciprocal speed of the body at rest", is postponed for a later paper.
[4] Feynman, Leighton, Sands. Feynman Lectures on Physics. Addison-Wesley Pub. Co.
[5] http://www.answers.com/topic/boltzmann-distribution
[6] Feynman, Leighton, Sands. Feynman Lectures on Physics. Addison-Wesley Pub. Co
[7] Md. Shah Alam and M.H. Ahsan - Mixed Number Lorentz Transformation. Vol.16.No.4
[8] Md. Shah Alam and M.H. Ahsan - Mixed Number Lorentz Transformation. Vol.16.No.4